\documentstyle[preprint,eqsecnum,aps]{revtex}
\begin{document}
\tightenlines
\draft
\title{The $U_A(1)$ problem and the role of correlated $q\bar{q}$\\ 
exchange in effective theories of QCD}
\author{M.R. Frank$^\ast $ and T. Meissner$^\dag $}
\address{
$\ast $Institute for Nuclear Theory, University of Washington, Seattle, 
WA 98195}
\address{$\dag $Department of Physics, Carnegie-Mellon University,
Pittsburgh, PA 15213}
\date{\today}
\maketitle
\begin{abstract}
The combined absence of physical realizations 
of the $U_A(1)$ symmetry possessed by the classical QCD action 
in the chiral limit, and of an isoscalar Goldstone 
boson associated with its spontaneous breakdown, has been dubbed the 
$U_A(1)$ problem.  A formal resolution of this problem proposed by 't Hooft 
relies on instantons to provide a mass to the would-be Goldstone 
boson ($\eta '$).  
An alternate scheme for the generation of an 
$\eta '$ mass proposed by Kogut and Susskind derives from quark 
annihilation into gluons and a strong infrared singularity in the gluon
propagator associated with confinement.  We demonstrate here how such 
diagrams are generated in quark based effective theories
by including a certain class of diagrams which arise from
correlated $q\bar{q}$ exchange and are of higher order    
$\frac{1}{N_c}$.
A low energy energy expansion of this corrections is of the form
discussed by Witten, di Vecchia and Veneziano.
\end{abstract}

\section{Introduction}
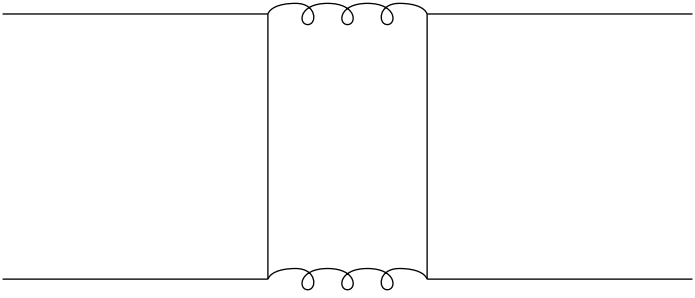
\begin{figure}
\caption{${\bar q}q$ annihilation diagram.}
\label{ff}
\end{figure}

The complexity of QCD makes the use of
effective theories a natural and important tool.  
The dominant role 
of dynamical chiral symmetry breaking at low energies 
and the mass gap in the hadron spectrum between the eight Goldstone modes 
($\pi$, $K$, and $\eta $) and the higher mass states has lead to the 
successful description of low-energy phenomena by chiral perturbation 
theory($\chi $PT) \cite{chipt}.  
In the case of the flavorsinglet pseudoscalar meson $\pi^0$
\footnote{We will ignore here the mixing between $\pi^0$, $\pi^3$
and $\pi^8$ because in the following we are working with massless current quarks.}
this division is not as clear, 
in particular with regard to the limit of a large number of colors $N_c$.  
In the large $N_c$ limit the $\pi^0$ is a Goldstone mode \cite{witten}, 
and its treatment 
by $\chi $PT as a higher mass state (which is integrated out) 
leads to the anomalous behavior of one of the chiral low energy 
(Gasser-Leutwyler) coefficients $L_7$, that is, $L_7\propto N_c^2$ \cite{holstein}.  
The extra factor of $N_c$  is supposed to result 
from the $N_c$-dependence of the 
$\pi^0$ mass, $m_{\pi^0}\propto 1/\sqrt{N_c}$, through the simple 
quark-antiquark annihilation picture shown in Fig.\ref{ff}.
Unfortunately things are not as simple as the diagram of Fig.\ref{ff} suggests.
It turns out that
the $\pi^0$ remains still massless in the chiral limit as long 
as the gluon propagators are the perturbative ones.
Kogut and Susskind \cite{ks} showed that using a nonperturbative ansatz
for the gluon propagator which diverges like 
$D(q^2) \propto \frac{1}{q^4}$ in the infrared limit $q^2 \to 0$
could actually generate a mass for the $\pi^0$.
What makes their approach even more attractive is the fact 
that such a gluon propagator generates a confining potential for the quarks
if used in a quark-quark interaction (for a review c.f. \cite{marciano}).
Furthermore it should be noted that such a $\frac{1}{q^4}$ singularity 
has also been extracted from studies of the gluon propagator in
the Dyson-Schwinger approach \cite{mandelstam,tuebingen}.

A formally different path to the $U_A(1)$ problem
was proceeded by 't Hooft \cite{thooft}
who proposed that $U_A (1)$ breaking as well as an $\pi^0$
mass occur from the interaction of the quarks
with a nontrivial topological vacuum configuration of the gauge field, 
called instantons.
Due to the large success of instanton based approaches 
to nonperturbative QCD (for a recent review c.f. \cite{shuryak})
many quark based effective chiral theories from the Nambu--Jona-Lasinio type
\cite{njl}
which use a contact quark-quark interaction
have incorporated this mechanism by adding a
't Hooft determinant which is a 6-quark point interaction in case
of $SU(3)$ flavor by hand to the original four quark
point interaction (for reviews c.f. \cite{njlreviews}).
On the other hand 
the formal solution proposed 
by 't Hooft \cite{thooft} seems to be in conflict with the simple $N_c$ 
counting arguments \cite{witten}.
In refs. \cite{witten,vecchia1,veneziano,vv,rosenzweig,nath} 
the general form of an effective chiral mesonic interaction 
which gives rise to a mass for the $\pi^0$ but not to the Goldstone
octet and is consistent with large $N_c$ QCD was derived and used
in various meson based effective chiral models.

It is the aim of this paper to
illustrate how the phenomenological notion proposed by 
Kogut and Susskind \cite{ks} arises in quark-based effective theories.
To do so we include explicitly 
correlated $q\bar{q}$ exchange diagrams from the non-Goldstone type.
In general 
such exchanges can include color octet correlations and are therefore 
not considered in meson-resonance \cite{chipt,gasser89} or 
dispersion-relation \cite{donoghue} extensions 
of chiral effective theories.  
Using a model gluon propagator with the Kogut-Susskind $\frac{1}{q^4}$
singularity and
performing a low energy expansion of these corrections in the same way
as it has been recently done for the leading order term \cite{frankmeissner} 
will then
render an effective interaction which is exactly from the form used in refs.
\cite{witten,vecchia1,veneziano,vv,rosenzweig,nath}.
Although our approach is only semiquantitative at the present stage
we are able to generate the $U_A (1)$ anomaly as well as an $\pi^0$ mass
directly in a quark based effective theory without introducing
explicitly instanton configurations.

Our paper is organized as follows:
In section 2 we exhibit the path from QCD to an effective quark-quark
interaction and its bosonization through the introduction of bilocal
fields.
These fields will be separated into Goldstone modes and other collective
${\bar q} q$ modes.
The main results of the saddle point expansion are briefly reviewed.
In section 3 we perform an integration over the non-Goldstone 
${\bar q} q$ modes which generates higher resonance exchange as well as
$\bar{q} q$ annihilation diagrams in addition  
to the saddle point effective action.
Among those we focus especially on the annihilation diagrams.
Section 4 introduces the technique for performing a low energy expansion of the
effective action which is generated by the annihilation diagrams.
We then discuss the properties of a model gluon propagator with a
Kogut-Susskind infrared $\frac{1}{q^4}$ singularity and show that such 
a singularity is sufficient to create a mass term for
the flavorsinglet Goldstone boson, i.e. the $\pi^0$,
which therefore breaks the $U_A (1)$ symmetry.
In section 5 we show that the low energy expansion of the 
effective action is of the form obtained in refs.
\cite{witten,vecchia1,veneziano,vv,rosenzweig,nath} which is consistent
with large $N_c$ QCD.
A summary and discussion is given in section 6.

\section{From QCD to the effective quark-quark interaction}
We begin by writing the QCD partition function in terms of an 
expansion in gluon $n$-point functions as 
\begin{eqnarray}
Z_{QCD}[\psi ,\bar{\eta },\eta ]
\equiv \int D\bar{q}DqDA \; && \mbox{exp}
\left( -S_{QCD}[\psi,\bar{q},q,A]+\bar{\eta }q+\bar{q}\eta \right) \nonumber \\
&& =\int D\bar{q}Dq \; \mbox{exp}
\left( -S[\psi,\bar{q},q]+\bar{\eta }q+\bar{q}\eta \right)\label{QCD}
\end{eqnarray}
with
\begin{equation}
S[\psi,\bar{q},q]\equiv \int \bar{q} 
\bigl( \not \! \partial _{x} +\psi (x)\bigr) q
+ \frac{1}{2}\int j_{\mu }^{a}(x)D_{\mu \nu }^{ab} (x-y)j_{\nu }^{b}(y) 
+ \frac{1}{3!}\int D^{abc}_{\mu \nu \lambda}
j_{\mu }^{a}j_{\nu }^{b}j_{\lambda }^{c}+ \cdots\label{npexp}
\end{equation}
where $j_{\nu }^{a}(x)\equiv \bar{q}(x)\frac{\lambda ^a}{2}\gamma _{\nu }q(x)$ 
is the quark color current, and $\psi (x)$ is an external source field.  The 
two-point function, $D_{\mu \nu }^{ab}$,   
for example is given by 
\begin{equation}
D_{\mu \nu }^{ab} (x-y)\equiv \int DA A^a_\mu(x)A^b_\nu(y)
\mbox{exp}\left( -S_{QCD}[0,0,0,A]\right).\label{dmn}
\end{equation}
It should be noted that this definition of the function $D_{\mu \nu }^{ab}$ 
does not include quark loops, which arise after the quark-field integration, 
and includes in principle all powers of $1/N_c$.  One should therefore, for the 
purposes of $N_c$ counting, write 
\begin{equation}
D_{\mu \nu }^{ab} (x-y)=\sum_{n=0}^{\infty}
[D^{(n)}]_{\mu \nu }^{ab} (x-y)\label{dnc}
\end{equation}
where $[D^{(n)}]_{\mu \nu }^{ab} \propto (1/N_c)^n$.  For example, 
$[D^{(0)}]_{\mu \nu }^{ab}$ contains only planar graphs.  

Here we will truncate the expansion (\ref{npexp}) to include only 
the two-point function, and thereby define an {\it effective} model based 
on the quark-quark interaction $D_{\mu \nu }^{ab} (x-y)$.  
This model truncation has become known has the global color model (GCM).
It retains all global symmetries of QCD, such as Poincare, chiral $U(3)\otimes U(3)$ and
global $SU(3)_c$ symmetry.
What is lost is the invariance under 
local $SU(3)_c$ gauge transformations and therefore
also renormalizability.
The inclusion of higher order terms in the series (\ref{npexp})
lies beyond our ability at the present time.
Nevertheless we can hope that by modeling the gluon 2 point function
$D_{\mu \nu }^{ab} (x-y)$ appropriately we are able to retain most of the important features 
of QCD, such as confinement, dynamical chiral symmetry breaking and
asymptotic freedom.
   
The GCM has been used
successfully for the description of low energy chiral physics
\cite{cahill1,roberts,gunner,frankmeissner}, meson form factors and 
spectra \cite{mesons,gunner}, the soliton 
\cite{soliton} and Fadeev \cite{cahill2} description of the
nucleon and QCD vacuum condensates \cite{meissner,km}.
Comprehensive reviews on this subject can be found in 
refs. \cite{robertswilliams} and \cite{tandyreviews}.

The partition function for this model is given by
\begin{equation} 
Z_{GCM}[\psi ,\bar{\eta },\eta ]\equiv \int D\bar{q}Dq \; \mbox{exp}
\left( -S_{GCM}[\psi,\bar{q},q]+\bar{\eta }q+\bar{q}\eta \right) 
\label{zgcm}
\end{equation} 
with
\begin{eqnarray}
S_{GCM}[\psi ,\bar{q},q]\equiv \int d^{4}xd^{4}y\Biggl\{ \bar{q}(x)\Bigl[ 
\bigl( \not \! \partial _{x} +\psi (x)&&  \bigr) \delta(x-y)\Bigr] 
q(y)\nonumber \\
 && + \frac{g^{2}}{2}j_{\nu 
}^{a}(x)D(x-y)j_{\nu }^{a}(y)\Biggr\} ,\label{gcm}
\end{eqnarray}
where for convenience a Feynman like gauge 
$D_{\mu \nu }^{ab} (x-y)=\delta^{ab} \delta _{\mu \nu } D(x-y)$ 
for the gluon propagator is employed. 

It should be noted that the Nambu--Jona-Lasinio(NJL) model \cite{njl}
is obtained from (\ref{gcm}) in the limit $D(x-y)=\delta (x-y)/{\cal M}^2$, 
with ${\cal M}$ the appropriate mass scale, rendering an effective four quark
contact interaction.
The introduction of an ultraviolet cutoff is required in this 
special case in order to make the quark loops finite.

As in our previous work, to make explicit contact with meson 
degrees of freedom we employ here the standard bosonization 
procedure \cite{kleinert,shrauner,cahill1}.  
This entails rewriting the partition 
function in terms of bilocal meson-like integration variables 
and expanding about the classical vacuum (saddle-point of the action).  

The resulting expression for the partition function in terms of the 
bilocal-field integration is $Z[\psi]={\cal N}\int D{\cal B}\; 
e^{-S[\psi,{\cal B}]}$ where the action is given by
\begin{equation}
S[\psi,{\cal B}]=-\mbox{TrLn}\left[ G^{-1}\right] +\int d^4xd^4y 
\frac{{\cal B}^{\theta 
}(x,y){\cal B}^{\theta }(y,x)}{2g^2D(x-y)},\label{2.13}
\end{equation}
and the quark inverse Green's function, $G^{-1}$, is defined as 
\begin{equation} 
G^{-1}(x,y)\equiv (\not \! \partial _x+\psi(x) )
\delta (x\! -\! y)+\Lambda ^{\theta }{\cal B}^{\theta }(x,y).\label{gf}
\end{equation}  
Here the quantity $\Lambda ^{\theta }$ 
is the direct product of Dirac, flavor $SU(3)$  and color matrices
\begin{equation}
\Lambda ^{\theta }=\frac{1}{2}\left( {\bf 1}_{D},i\gamma 
_{5},\frac{i}{\sqrt{2}}\gamma _{\nu },\frac{i}{\sqrt{2}}\gamma _{\nu 
}\gamma _{5}\right) \otimes \left( \frac{1}{\sqrt{3}}{\bf 
1}_{F},\frac{1}{\sqrt{2}}\lambda _F^a\right) \otimes \left( \frac{4}{3}{\bf 
1}_{C},\frac{i}{\sqrt{3}}\lambda _C^{a}\right) \label{fierz1}
\end{equation} 
which arises from Fierz reordering the current-current interaction in
(\ref{gcm}) 
\begin{equation}
{\Lambda^{\theta}}_{ji}  
{\Lambda^{\theta}}_{lk}  
=
\left ( \gamma_\mu \frac{\lambda^a}{2} \right )_{jk} 
\left ( \gamma_\mu \frac{\lambda^a}{2} \right )_{li} \, . 
\label{fierz2}
\end{equation}

The saddle-point of the action is defined as
$\left. \frac{\delta S}{\delta {\cal B}}\right| _{{\cal B}_0,\psi =0}=0$ and 
is given by
\begin{equation}
{\cal B}^{\theta }_0(x-y)=g^2D(x-y)\mbox{tr}\left[ \Lambda 
^{\theta }G_{0}(x-y)\right] 
.\label{spoint}
\end{equation}
These configurations 
provide self-energy dressing of the quarks through the definition 
$\Sigma (p)\equiv \Lambda ^{\theta }{\cal B}^{\theta }_0(p)=
i\not \! p\left[ A(p^2)-1\right] +B(p^2)$, where  
\begin{equation}
\left[ A(p^{2})-1\right] p^{2} =g^{2} \frac{8}{3}\int \frac{d^{4}q}{(2\pi 
)^{4}}D(p-q)\frac{A(q^{2})q\cdot p}{q^{2}A^{2}(q^{2})+B^{2}(q^{2})}, 
\label{aeq}
\end{equation}
and
\begin{equation}
B(p^{2})=g^{2}\frac{16}{3}\int \frac{d^{4}q}{(2\pi 
)^{4}}D(p-q)\frac{B(q^{2})}{q^{2}A^{2}(q^{2})+B^{2}(q^{2})} 
.\label{beq}
\end{equation}
This dressing comprises the 
notion of ``constituent'' quarks by providing a mass
$M(p^2)=B(p^2)/A(p^2)$.

The bilocal fields are then expanded about the saddle point as, 
\begin{equation}
{\cal B}^{\theta }(x,y)=
{\cal B}^{\theta }_0(x-y)+\hat{\cal B}^{\theta }(x,y), \label{fluct}
\end{equation}
to generate the effective interactions of 
the fluctuations, $\hat{\cal B}$.  At this point these interactions 
are produced by 
the quark-field determinant $\mbox{TrLn}\left( \not \! \partial + 
\Sigma +\Lambda ^{\theta }\hat{\cal B}^{\theta }\right) $ and therefore 
occur through the dressed-quark loops.  

The connection between the bilocal fluctuation fields and the local fields 
of standard hadronic field-theory phenomenology is made as follows.  The 
bilocal fields contain information about internal excitations of the 
$q\bar{q}$ pair in addition to their net collective or center-of-mass motion 
which is to be associated with the usual local field variables.  A separation 
of the internal and center-of-mass dynamics is achieved by considering the 
normal modes of the free kinetic operator of the bilocal fields in a manner 
which is analogous to the interaction representation of standard quantum 
field theory.  
Details of the localization procedure 
can be found in refs. \cite{kleinert,shrauner,cahill1,tandyreviews}.
The process amounts to a projection of the 
bilocal field $\hat{\cal B}^{\theta }$ onto a complete set of internal 
excitations $\Gamma _{n}^{\theta }$ with the 
remaining center-of-mass degree of freedom represented by the coefficients 
$\pi ^{\theta }_{n}(P)\equiv \int d^4q\hat{\cal B}^{\theta }(P,q)
\Gamma _{n}^{\theta }(P,q)$.  The bilocal fluctuations can thus be written as 
\begin{equation}
\hat{\cal B}^{\theta }(P,q)=\sum _n\pi ^{\theta }_{n}(P)
\Gamma _{n}^{\theta }(P,q).
\label{lfields}
\end{equation}

The functions $\Gamma _{n}^{\theta }$ are in general eigenfunctions of the 
the free kinetic operator of the bilocal fields.  At the mass shell point, 
$P^2=-M_n^2$, they satisfy the homogeneous Bethe-Salpeter equation 
in the ladder approximation for the 
given quantum numbers $\theta $ and mode $n$.  This modal expansion is 
then used to localize the action.  

Our previous work \cite{frankmeissner} was restricted to the study 
of the Goldstone 
octet by explicitly neglecting all other fluctuations.  
Here we wish to 
extend that work by including and explicitly integrating out all of the 
$q\bar{q}$ fluctuations 
except the would be Goldstone ones, i.e. the 
ground state pseudoscalar nonet.  In doing so 
we will generate resonance-exchange and quark-annihilation diagrams; the 
latter being associated with the $1/N_c$ corrections.  To this end we 
write the full bilocal field of (\ref{fluct}) as 
\begin{equation}
{\cal B}^{\theta }(x,y)={\cal B}_U^{\theta }(x,y)+
{\cal B}_E^{\theta }(x,y)\label{bfield}
\end{equation}
where ${\cal B}_U^{\theta }(x,y)$ includes the vacuum configuration and 
the ground-state pseudoscalar nonet fluctuations, 
\begin{equation}
\Lambda ^{\theta }{\cal B}_U^{\theta }(x,y)\equiv \Sigma (x-y) + 
B(x-y)\left[ U_5\left(\frac{x+y}{2}\right) -1\right] ,\label{bgfield}
\end{equation}
and ${\cal B}_E^{\theta }(x,y)$ includes all other excitations, 
\begin{equation}
{\cal B}_E^{\theta }(P,q)\equiv \sum ^\prime_n\pi ^{\theta }_{n}(P)
\Gamma _{n}^{\theta }(P,q)\label{befield}.
\end{equation}
It is important to note that in (\ref{befield}) the sum is restricted, 
which has been denoted by the $\prime$. 
Therefore, due to the completeness of the $\Gamma _{n}^{\theta }(P,q)$, 
${\cal B}_E^{\theta }(x,y)$ is proportional to a projection operator that 
excludes the ground-state pseudoscalar nonet fluctuations.  Further in 
(\ref{bgfield}) $U_5(x)$ is defined as 
\begin{eqnarray}
U_5(x)&\equiv &P_RU(x)+P_LU^{\dag}(x)\nonumber \\
U(x)&\equiv &e^{i\pi ^0/f_\pi}e^{i\lambda ^a\pi ^a/f_\pi}\label{u5},
\end{eqnarray}
with $P_{R,L}$ the usual chiral projection operators.

In terms of the would be Goldstone fields ${\cal B}_U^{\theta }(x,y)$ and the
excitations ${\cal B}_E^{\theta }(x,y)$ the action of (\ref{2.13}) 
is given by 
\begin{eqnarray}
{\cal S}=-&&\mbox{TrLn}\left[ G_U^{-1}\right] 
-\mbox{TrLn}\left[ 1+G_U\Lambda^\theta{\cal B}_E^{\theta }\right] \nonumber \\
&&+\int d^4xd^4y \frac{{\cal B}_E^{\theta 
}(x,y){\cal B}_E^{\theta }(y,x)}{2g^2D(x-y)}
+\int d^4xd^4y \frac{{\cal B}_U^{\theta 
}(x,y){\cal B}_E^{\theta }(y,x)}{g^2D(x-y)}\label{ueaction},
\end{eqnarray}
where we have discarded a constant term.  
Expanding (\ref{ueaction}) to second order in the 
excitations ${\cal B}_E^{\theta }$ gives
\begin{eqnarray}
{\cal S} = -\mbox{TrLn}\left[ G_U^{-1}\right] 
 &+&
\int d^4xd^4y {\cal J}^{\theta }(x,y){\cal B}_E^{\theta}(y,x)
\nonumber \\
 &+&
\frac{1}{2}\int d^4xd^4yd^4x'd^4y'\; {\cal B}_E^{\theta }(x,y)
{\cal D}^{-1}_{\theta \theta'}(x,y;x',y')
{\cal B}_E^{\theta '}(y',x')
\label{eq219} 
\end{eqnarray}
where 
\begin{equation}
{\cal D}^{-1}_{\theta \theta'}(x,y;x',y')\equiv \frac{\delta_{\theta \theta'}
\delta(x-x')\delta(y-y')}{g^2D(x-y)}+\mbox{tr}\left[ G_U(x',x)
\Lambda^\theta G_U(y,y')\Lambda^{\theta '}\right] \label{D-1}
\end{equation}
and 
\begin{equation}
{\cal J}^{\theta }(x,y)\equiv \frac{{\cal B}_U^{\theta }(x,y)}{g^2D(x-y)}
-\mbox{tr}\left[ G_U(x,y)\Lambda^\theta\right] .\label{jtheta}
\end{equation}
In the first term of (\ref{eq219}) the fluctuations are 
restricted to the pseudoscalar (would be Goldstone boson) nonet.
As we will discuss in detail 
in the next section the second and the third term of  (\ref{eq219})
will give rise to higher resonance exchange diagrams 
as well as ${\bar q} q$
annihilation diagrams.

So far only the first term of (\ref{eq219}) has been studied.
Its real part can be written as
\footnote{In the following we will restrict ourselves only
to the real part of the action.
The imaginary part $\Im [ {{\cal S}} ]$ gives rise
to the Wess-Zumino term after gradient expansion which 
is connected to chiral (non abelian) anomaly.} 
\begin{equation}
{\cal S}_U \equiv \Re \left [ {\cal S} 
\{ {\cal B}_E^{\theta } =0 \} \right ] = 
-\frac{1}{2}\mbox{TrLn}
\left( G_U^{-1}\left[ G_U^{-1}\right] ^{\dag }\right) ,\label{res}
\end{equation}
where $G_U^{-1}$ is, from (\ref{gf}) and (\ref{bgfield}), given by 
\begin{equation}
G_U^{-1}(x,y)=\gamma \cdot \partial _xA(x-y)+\psi \left( 
\frac{x+y}{2}\right) \delta(x-y) +B(x-y)U_5\left( 
\frac{x+y}{2}\right) .\label{gfu}
\end{equation}

The chiral effective action is obtained by performing a gradient expansion of 
(\ref{res})(details are given in ref. \cite{frankmeissner}).  
The result to fourth order is (in Euclidean space)
\begin{eqnarray}
{\cal S}_U
=\int d^4x &\Biggl\{ & \frac{f^2_{\pi }}{4}\mbox{tr}_F
\left[ (\partial _{\mu}U)(\partial _{\mu}U^{\dag})\right] 
-\frac{f^2_{\pi }}{4}
\mbox{tr}_F \left[ U\chi ^{\dag} +\chi U^{\dag}\right] 
\nonumber \\
&-&L_1\left( \mbox{tr}_F \left[ (\partial _{\mu}U)
(\partial _{\mu}U^{\dag})\right] \right) ^2 -L_2
\mbox{tr}_F \left[ (\partial _{\mu}U)(\partial _{\nu}U^{\dag})\right]
\cdot \mbox{tr}_F \left[ (\partial _{\mu}U)(\partial _{\nu}
U^{\dag})\right] \label{chiact} \\
&-&L_3\mbox{tr}_F \left[ (\partial _{\mu}U)(\partial _{\mu}U^{\dag})
(\partial _{\nu}U)(\partial _{\nu}U^{\dag})\right] +L_5
\mbox{tr}_F \left[ (\partial _{\mu}U)(\partial _{\mu}U^{\dag})
(U\chi ^{\dag} +\chi U^{\dag})\right] \nonumber \\
&-&L_8\mbox{tr}_F \left[ \chi U^{\dag}\chi U^{\dag} + 
U\chi ^{\dag }U\chi ^{\dag }\right] \Biggr\} ,\nonumber
\end{eqnarray}
where $\chi (x)=-2\langle \bar{q}q\rangle\psi (x)/f^2_{\pi }$ and the 
remaining trace is over flavor.  
In obtaining this result the equation of motion 
\begin{equation}
(\partial ^2U)U^{\dag}+(\partial _{\mu}U)(\partial _{\mu}U^{\dag})
+\frac{1}{2}(\chi U^{\dag }-U\chi ^{\dag })=0\label{eqmo}
\end{equation}
and the $SU(3)$ relation \cite{holstein}
\begin{eqnarray}
&{\mbox{tr}_F}& 
\left[ (\partial _{\mu}U)(\partial _{\nu}U^{\dag})
(\partial _{\mu}U)(\partial _{\nu}U^{\dag})\right] = 
\nonumber \\
\frac{1}{2} \Biggl (
&{\mbox{tr}_F}&  \left[ (\partial _{\mu}U)
(\partial _{\mu}U^{\dag}) \right ] \Biggr ) ^2 
\nonumber \\
 + \, &{\mbox{tr}_F}& \, 
\left[ (\partial _{\mu}U)(\partial _{\nu}U^{\dag})\right]
\cdot \mbox{tr}_F \left[ (\partial _{\mu}U)(\partial _{\nu}
U^{\dag})\right] 
\, - \, 2 \, {\mbox{tr}_F} \left[ (\partial _{\mu}U)
(\partial _{\mu}U^{\dag})
(\partial _{\nu}U)(\partial _{\nu}U^{\dag} )
\right ] 
\label{su3}
\end{eqnarray}
have been used.  
Explicit forms of the coefficients are also 
given in ref. \cite{frankmeissner}.    

Finally it should be noted that in terms of $N_c$ counting ${\cal S}_U$
is of order ${\cal O}(N_c)$.

\section{Integration over the non-Goldstone modes; $\frac{1}{N_c}$ Corrections
to the effective action}
\subsection{Integration over the Non-Goldstone Modes}
As the next step we now perform the functional
integration over the excitation fields
$ {\cal B}_E^{\theta }$ but not over the would be Goldstone fields 
$ {\cal B}_G^{\theta }$, or, in other words, we 
treat the would be Goldstone fields at the 
mean field level (stationary phase approximation).
It is convenient to introduce the abbreviations:
\begin{eqnarray}
{\cal D}^{-1}_{\theta \theta'}(x,y;x',y') &\equiv&
\langle xy\theta \vert {\cal D}^{-1} \vert x' y' \theta' \rangle \nonumber \\
 {\cal B}_G^{\theta } (x,y) &\equiv&
 \langle {\cal B}_G \vert xy \theta \rangle \nonumber \\
 {\cal B}_E^{\theta } (x,y) &\equiv&
 \langle {\cal B}_E \vert xy \theta \rangle \nonumber \\
{\cal J}^{\theta }(x,y) &\equiv& 
\langle {\cal J} \vert xy \theta \rangle
\label{eq31}
\end{eqnarray}
so that (\ref{eq219}) reads
\begin{equation}
{\cal S} = {\cal S}_U
+ 
\frac{1}{2} \langle {\cal B}_E \vert {\cal D}^{-1} \vert {\cal B}_E \rangle
+
\langle {\cal J} \vert {\cal B}_E \rangle .
\label{eq32}
\end{equation}
We now apply the general rule for the functional integration
over a boson field $\phi$
\begin{equation}
\int D \phi e^{-\frac{1}{2} \int \phi {\cal M}^{-1} \phi + j \phi }
= \left ( \mbox{Det} {\cal M}^{-1} \right )^{-\frac{1}{2}} 
e^{-\frac{1}{2} \int j {\cal M} j }    
\label{eq33}
\end{equation}
from which we obtain the complete mean field effective action 
for the would be Goldstone modes as:
\begin{equation} 
{\cal S} = {\cal S}_U + {\cal S}_{res} + {\cal S}_{ann}
\label{eq34}
\end{equation}
where ${\cal S}_U$ has been discussed in the last section and
the two other terms are given by 
\begin{eqnarray}
{\cal S}_{res} &=& \frac{1}{2} \langle {\cal J} \vert {\cal D} \vert {\cal J}
\rangle ,  \label{eq35} \\
{\cal S}_{ann} &=& - \frac{1}{2} \mbox{TrLn} \left [ {\cal D}^{-1} \right ] .
\label{eq36}
\end{eqnarray}
Let us now discuss these new two terms,
which arise due to the inclusion of correlated ${\bar q}q$ exchange
of the non-Goldstone type, in more detail.

\subsection{Resonance Exchange Diagrams} 
From the form of (\ref{jtheta}) 
and the saddle point condition (\ref{spoint})
it follows that ${\cal J}^\theta$ 
has to be at least of first order in the would be Goldstone fields
$U_5 -1$. 
Furthermore it should be recalled that the field 
${\cal B}_E^{\theta}$ defined in (\ref{befield}) is proportional 
to a projection operator that excludes all the would be Goldstone modes.
Therefore the first contribution of ${\cal J}^\theta$ to 
the second term in (\ref{eq219}) must be of second order in the
would be Goldstone fields.
Altogether it follows that ${\cal S}_{res}$ in (\ref{eq35}) 
must be at least of
4th order in the would be Goldstone fields.

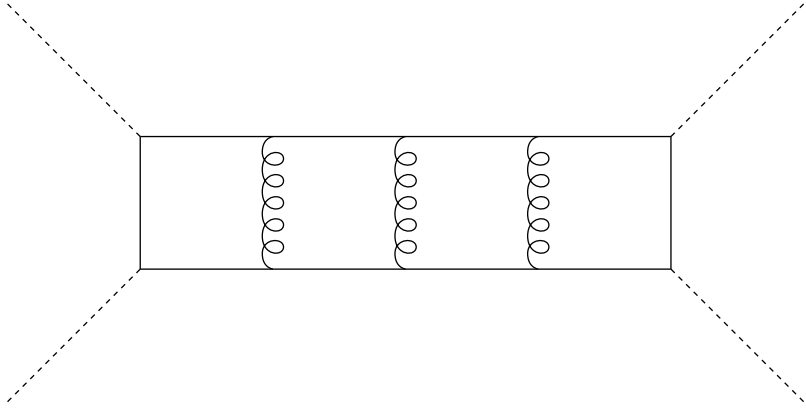
\begin{figure}
\caption{Typical diagram appearing in ${\cal S}_{res}$.
The dashed lines denote external would be Goldstone
background fields $U_5 -1$, whose minimum number is 4.
The diagram describes therefore an interaction between would be
Goldstone bosons (e.g. $\pi -\pi$ scattering). 
These diagrams are of ${\cal O}(N_c)$.}
\label{fig1}
\end{figure}

To be more explicit let us separate the inverse quark propagator
${G_U}^{-1}$ 
in (\ref{gfu}) in case of a vanishing external scalar source $\psi \equiv 0$ 
into a ``free'' part and chiral field ``background''
\begin{equation}
{G_U}^{-1} (x,y) = {G_0}^{-1} (x-y) + B(x-y) \left ( U_5 \left (
\frac{x+y}{2} \right ) - 1 \right )
\label{eq37}
\end{equation}
with
\begin{equation}  
{G_0}^{-1} (x-y) \equiv \gamma \partial_x A(x-y) + B(x-y) 
\label{eq38}
\end{equation} 
which gives rise to an expansion of $G_U$ in terms of $B(U_5 -1)$
\begin{equation}
G_U = G_0 - G_0 \left [B(U_5 -1) \right ] G_0 
+ G_0  \left [ B(U_5 -1) \right ] G_0  \left [ B(U_5 -1) \right ] G_0 
\pm \dots \, .
\label{eq39}
\end{equation}
A similar expansion may be obtained for the operator 
${\cal D}$ which was defined in (\ref{D-1}) and describes the propagation
of a resonance (${\bar q} q$ excitation) in the presence of a 
``background'' $U_5 - 1$:
\begin{eqnarray}
{\cal D}^{-1} &=& {\cal D}_0 ^{-1} + {\cal V} \nonumber \\
{\cal D} &=& {\cal D}_0 - {\cal D}_0 {\cal V} {\cal D}_0 
+ {\cal D}_0 {\cal V} {\cal D}_0  {\cal V} {\cal D}_0
\pm 
\label{eq310}
\end{eqnarray} 
where
\begin{eqnarray}
\left [ {{{\cal D}_0}^{-1}}\right ]_{\theta \theta'} (x,y;x',y')
&=& \frac{\delta_{\theta \theta'}
\delta(x-x')\delta(y-y')}{g^2D(x-y)} \nonumber \\
{\cal V}_{\theta \theta'}(x,y;x',y') &=& 
 \mbox{tr}\left[ G_U(x',x)
\Lambda^\theta G_U(y,y')\Lambda^{\theta '}\right ] .
\label{eq311}
\end{eqnarray}
Inserting these expansion into the definition of ${\cal S}_{res}$
(\ref{eq35}) together with (\ref{jtheta}) 
and reapplying the Fierz relation (\ref{fierz2})
renders an  
expansion for ${\cal S}_{res} $.
A typical diagram is displayed in Fig.\ref{fig1}.
The dashed lines stand for $U_5 -1$ ``background'' field insertions,
whose minimal number are four, as we have discussed above.

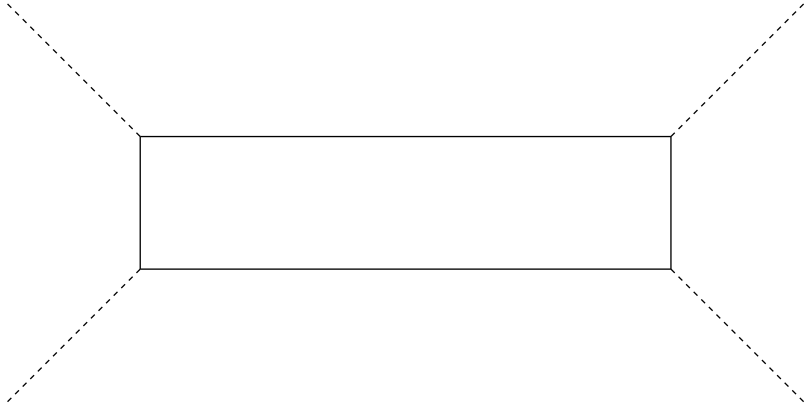
\begin{figure}
\caption{The same as Fig.{\protect{\ref{fig1}}} only with a simple
(dressed) ${\bar q} q$ loop. These diagrams come from ${\cal S}_U$
and are of ${\cal O}(N_c)$ as well.}
\label{fig2}
\end{figure}

All those diagrams are of order ${\cal O}(N_c)$ such as 
${\cal S}_U $ itself. 
As it has already been discussed in some more detail
in ref. \cite{frankmeissner} 
they generate higher resonance exchange contributions to the 
interactions between the would be Goldstone bosons
in addition to simple quark loop diagrams such as Fig.
\ref{fig2} coming from ${\cal S}_U$. 
An example would be e.g.
the $\rho$ meson exchange pole in $\pi - \pi$ scattering.
However, as also discussed in ref. \cite{frankmeissner},
the main contribution to the chiral low energy coefficients
$L_i$ is already given by the simple quark loop diagrams 
Fig.\ref{fig2}, which obviously can reproduce the low energy tail
of the resonances satisfactorily.
Therefore the diagrams arising from
the new action ${\cal S}_{res}$ (Fig.\ref{fig1}) do not seem to play
a very important role for low energy chiral dynamics. 

\subsection{${\bar q}q$ Annihilation Diagrams}
Let us now come to the main focus of this paper, which are 
the ${\bar q}q$ annihilation diagrams.
They arise from the action ${\cal S}_{ann}$
(\ref{eq36}).
In order to see this we use the expansion (\ref{eq310}) in 
(\ref{eq36}) and obtain after dropping a constant 
\begin{eqnarray}
{\cal S}_{ann} &=& -\frac{1}{2} \mbox{TrLn} \left [{\cal D}_0 ^{-1} \right ]
- \frac{1}{2} \mbox{TrLn} \left [ \openone + {\cal D}_0 {\cal V} \right ]
\nonumber \\
&\equiv& - \frac{1}{2} \mbox{Tr} \left [  {\cal D}_0 {\cal V} \right ]
+ \frac{1}{4} \mbox{Tr} \left [  {\cal D}_0 {\cal V} {\cal D}_0 {\cal V} 
\right ] 
- \frac{1}{6} \mbox{Tr} 
\left [  {\cal D}_0 {\cal V} {\cal D}_0 {\cal V}  {\cal D}_0 {\cal V} \right ] 
\pm 
\dots \, .
\label{eq312}
\end{eqnarray}
Using the explicit definitions (\ref{eq311}) and again reapplying the
Fierz relation (\ref{fierz2}) leaves us with
\begin{eqnarray}
&& {\cal S}^{1g} \, = \, 
- \frac{1}{2}  \mbox{Tr} \left [  {\cal D}_0 {\cal V} \right ] =
\nonumber  \\
&& \int d^4 x_1 d^4 x_2 g^2 D(x_1 - x_2) 
\mbox{tr} \left [ G_U (x_1,x_1) \gamma_\mu \frac{\lambda^a}{2} \right ]
\mbox{tr} \left [ G_U (x_2,x_2) \gamma_\mu \frac{\lambda^a}{2} \right ] ,
\label{eq312a}
\end{eqnarray}
diagrammatically shown in Fig.\ref{fig3}(a), which vanishes due to the color 
trace.
The first non vanishing contribution is
the 2 gluon annihilation diagram (Fig.\ref{fig3}(b))
\begin{eqnarray}
&&{\cal S}^{2g} \, = \,  {\frac{1}{4}} 
\mbox{Tr} \left [  {\cal D}_0 {\cal V} {\cal D}_0 {\cal V} 
\right ]
\, = \, 
\nonumber \\
&&
{\frac{1}{4}} \int d^4 x_1 d^4 x_2 d^4 y_1 d^4 y_2 \,
{ g^2} D(x_1 - y_1) {g^2} D(x_2  - y_2)
\nonumber \\
&& 
\mbox{tr} \left [ G_U (x_2,x_1) \gamma_{\mu_1} \frac{\lambda^{a_1}}{2}
G_U (x_1,x_2) \gamma_{\mu_2} \frac{\lambda^{a_2}}{2}
\right ]
\nonumber \\
&&
\mbox{tr} \left [ G_U (y_1,y_2) \gamma_{\mu_2} \frac{\lambda^{a_2}}{2}
 G_U (y_2,y_1) \gamma_{\mu_1} \frac{\lambda^{a_1}}{2}
\right ] \, .
\label{eq313}
\end{eqnarray}
The 3 gluon annihilation diagram (Fig.\ref{fig3}(c)) gives
\begin{eqnarray}
&& {\cal S}^{3g} \, = \,  - {\frac{1}{6}} 
\mbox{Tr} \left [  {\cal D}_0 {\cal V} {\cal D}_0 {\cal V} {\cal D}_0 {\cal V}
\right ]
\, = \, 
\nonumber \\
&& - {\frac{1}{6}} \int d^4 x_1 d^4 x_2 d^4 x_3 d^4 y_1 d^4 y_2 d^4 y_3 \,
{ g^2} D(x_1 - y_1) {g^2} D(x_2  - y_2) {g^2} D(x_3  - y_3)
\nonumber \\
&& 
\mbox{tr} \left [ 
G_U (x_1,x_3) \gamma_{\mu_3} \frac{\lambda^{a_3}}{2}
G_U (x_3,x_2) \gamma_{\mu_2} \frac{\lambda^{a_2}}{2}
G_U (x_2,x_1) \gamma_{\mu_1} \frac{\lambda^{a_1}}{2}
\right ] 
\nonumber \\ &&
\mbox{tr} \left [ 
\gamma_{\mu_1} \frac{\lambda^{a_1}}{2} G_U (y_1,y_2) 
\gamma_{\mu_2} \frac{\lambda^{a_2}}{2} G_U (y_2,y_3)
\gamma_{\mu_3} \frac{\lambda^{a_3}}{2} G_U (y_3,y_1)
\right ]
\label{eq313a}
\end{eqnarray} 
and so forth.
From the diagrams of Fig.\ref{fig3} 
we conclude that ${\cal S}_{ann}$
is of one order $\frac{1}{N_c}$ higher than ${\cal S}_U$
as well as ${\cal S}_{res}$.  

%

\section{Low Energy Expansion; Mass terms for the pseudoscalar nonet}
The presence of ${\bar q}q$ annihilation diagrams (Fig.\ref{fig3})
which arise from the integration over higher mass states,
does not guarantee a priori a 
finite mass 
for the flavorsinglet pseudoscalar would be Goldstone boson, the $\pi^0$.
This point has been commonly dubbed the {\it $U_A (1)$ problem}
in the literature.
In order to understand this point in the context of our approach
let us investigate the low energy structure of the effective
action ${\cal S}_{ann} $ in some more detail.

\subsection{Gradient Expansion of the Quark Propagator $G_U$ in the Presence
of an External Background Field $U$}
The first step to do this is to expand the quark propagator $G_U (x_1,x_2)$
in the presence of an external chiral field $U_5$ in terms of the
momenta of this external field.
It is convenient to write the matrix $U_5$ as 
\begin{eqnarray}
U_5 &=& e^{i\gamma_5 \frac{\pi^i}{f_\pi} \lambda^i} \equiv {g_5}^2 \nonumber \\
g_5 &=& e^{i\gamma_5 \frac{\pi^i}{f_\pi} \frac{\lambda^i}{2} } ,
\label{eq41}
\end{eqnarray}
where the index $i$ runs from $0$ to $8$, i.e. includes the full pseudoscalar nonet.
First of all we observe that the inverse quark propagator ${G_U}^{-1}$ defined in
eq.(\ref{eq37}) can be expanded as:
\begin{eqnarray}
&& <x_1 | {G_U}^{-1} | x_2 > = {G_U}^{-1} (x_1,x_2) =
\nonumber \\
&& 
g_5 \left ( \frac{x_1 + x_2}{2} \right ) \;
{G_0}^{-1} (x_1 - x_2) \;
g_5 \left ( \frac{x_1 + x_2}{2} \right ) \; 
=
\nonumber \\
&&
g_5 (x_1) \,
{G_0}^{-1} (x_1 - x_2) \, 
g_5 (x_2)
\nonumber \\
&&
+ \frac{1}{2} (x_2 - x_1)_\mu 
\left \{ \partial_\mu g_5 (x_1) 
{G_0}^{-1} (x_1 - x_2)  g_5 (x_2) 
- g_5 (x_1) 
{G_0}^{-1} (x_1 - x_2)  \partial_\mu g_5 (x_2) \right \} + \dots
\label{eqgum1} \, ,
\end{eqnarray}
which in turn gives rise to an expansion
\begin{equation}
<x_1 | {G_U} | x_2 > = {G_U} (x_1,x_2) =
{g_5}^{\dag} (x_1) \; \{ G_0 (x_1-x_2) + G_1(x_1,x_2) + \dots \} \; {g_5}^{\dag} (x_2)
\label{gexp}
\end{equation}
where
\begin{eqnarray}
&&  G_1(x_1,x_2) = 
<x_1 | G_1 | x_1 > = 
\nonumber \\
&& 
- \frac{1}{2} \int d^4 y d^4 z  (z-y)_\mu 
G_0(x_1,y) \, \cdot 
\nonumber \\
&&
\left [ {g_5}^{\dag} (y) \partial_\mu g_5 (y)   {G_0}^{-1} (y,z) -
{G_0}^{-1}  (y,z) \partial_\mu g_5 (z) {g_5}^{\dag} (z)
\right ] \, \cdot \,  
G_0(z,x_2) 
\label{eqg1}
\end{eqnarray}
or in momentum space
\begin{eqnarray}
&&  G_1(x_1,x_2) = 
<x_1 | G_1 | x_1 > = 
+ \frac{1}{4}
\int d^4 R 
\int \frac{d^4 k}{(2\pi)^4} \frac{d^4 q}{(2\pi)^4} 
e^{iq \frac{x_1 + x_2 }{2}}
e^{i k (x_1 - x_2)}
e^{ i R q}
\nonumber \\
&& \frac{(-i)(\not{\! k} + \frac{1}{2} \not{\! q}) A (k + \frac{1}{2} q) +
B(k + \frac{1}{2} q) } { X (k + \frac{1}{2} q) } \, \cdot    
\nonumber \\
&&
\Biggl \{
\Bigl [
\left (
{U_5}^\dag (R) \partial_\mu U_5 (R) -  U_5 (R) \partial_\mu {U_5}^\dag (R)
\right )
(- )
\left ( \gamma_\mu A (k) + 2 \not{\! k} k_\mu A^\prime (k) 
\right )
\Bigr ]
\nonumber \\
&&
+ 
\Bigl [
\partial_\nu
\left (
{U_5}^\dag (R) \partial_\mu U_5 (R) +  U_5 (R) \partial_\mu {U_5}^\dag (R)
\right ) \, \cdot
\nonumber \\
&& (-i)
\left (
(\gamma_\mu k_\nu + \gamma_\nu k_\mu + \delta_{\mu\nu} \not{\! k} )
A^\prime (k)
+ 2 \not{\! k} k_\mu k_\nu A^{\prime\prime} (k)
\right )
\Bigr ]
\nonumber \\
&&
+
\Bigl [
\partial_\nu
\left (
{U_5}^\dag (R) \partial_\mu U_5 (R)
\right )
(-2)
\left ( 
\delta_{\mu\nu} B^\prime (k) + 2  k_\mu k_\nu B^{\prime\prime} (k)
\right ) 
\Bigr ] \; + \; \dots \Biggr \}  \cdot \,
\nonumber \\
&& \frac{(-i)(\not{\! k} - \frac{1}{2} \not{\! q}) A (k - \frac{1}{2} q) +
B(k - \frac{1}{2} q) } { X (k - \frac{1}{2} q) } \, , 
\label{fourier}
\end{eqnarray}
where $X(s) \equiv s A^2(s) + B^2(s)$ and the $\dots$ denote higher gradients
in the chiral field.

We are now ready to study the low energy expansion of the action
${\cal S}_{ann}$ by putting the expansion (\ref{gexp}) into the 
series for  the multigluon annihilation diagrams (\ref{eq312a},\ref{eq313},\ref{eq313a},...).
Because of 
\begin{equation} 
{g_5}^{\dag} \gamma_\mu {g_5}^{\dag} = \gamma_\mu
\label{eq43}
\end{equation}
the lowest order in the expansion (\ref{gexp}) will 
for any number of gluon annihilations 
only render terms which are independent 
of the chiral field $U_5$, whereas the higher order terms contain at least one
gradient.
One therefore could naively conclude that it is impossible to generate a mass term for
the flavorsinglett $\pi^0$, which must be proportional to $({\pi^0})^2$.
This of course just reflects the fact that all the terms in
(\ref{eq312a},\ref{eq313},\ref{eq313a},...)
and therefore ${\cal S}_{ann}$    
should be invariant under $U_A (1)$ rotations
as long as one was dealing merely with finite matrices.
This is however not the case.
The traces $\mbox{Tr}$ are all infinite operator traces
and the nonlocality in the gluon 2 point functions $D(x_i - y_i)$  
could produce infrared singularities.
It is therefore necessary to study the IR singularity structure 
of the gluon propagator $D$ in more detail.

\subsection{Infrared Behavior for a Model Gluon Propagator with a Kogut-Susskind Singularity}
In order to do so
let us consider a model gluon two point function which we write as
\begin{equation}
g^2 D(p^2) = g^2 D_s (p^2) +  g^2 D_r (p^2)
\label{eqdsplit}
\end{equation}
For the IR singular part we consider a
Kogut-Susskind type form,
which has also been used by Marciano and Pagels \cite{marciano}
\begin{equation}
 g^2 D_s (p^2)
= \left ( \frac{\mu^2}{p^2} \right ) ^{1-\epsilon} \frac{1}{p^2} 
\, ,
\label{eq48}
\end{equation}
where $\mu$ is an energy scale and $\epsilon$ an infrared regulator and regarded as small.
In this context we want to remind that in the so called Abelian approximation,
which consists in neglecting ghosts, the gluon 2 point function $D(p^2)$ is related to the 
strong running coupling $\alpha_s (p^2)$ through
\begin{equation}
g^2 D(s) = \frac{4\pi \alpha_s (s)}{s} \, .
\label{alphas}
\end{equation}
For the gluon propagator in (\ref{eq48}) this would imply that $\alpha_s (s) \propto \frac{1}{s}$
which is consistent with infrared slavery.
More discussion on this subject can be found in \cite{robertswilliams,tandyreviews} and references
therein.
For the regular part we can assume asymptotic scaling in the UV region, i.e. 
\begin{equation}
{D_r (s)} \propto \frac{1}{s \mbox{Ln}(\frac{s}{{\Lambda_{QCD}}^2} ) } 
\label{asy}
\end{equation}
for $s\to\infty$,
as it was done in most of the previous studies (c.f. \cite{robertswilliams,tandyreviews} and refs. therein).
As we will see shortly our conclusions will not depend on the details of the form for 
${D_r (s)}$ as long as it is less singular than $1/s^2$ in the IR.
The solutions of the Dyson-Schwinger equations (\ref{aeq}) and (\ref{beq}) are in the limit $\epsilon \to 0$
\begin{equation}
B(s) \, \propto \, \sqrt{\frac{\mu^2}{3\epsilon \pi^2} - 4s} \, ,
\label{eq49}
\end{equation}
whereas $A(s)$ as well as all derivatives $B^\prime (s) , A^\prime (s), \dots$ remain finite.
For large values of $s$ the solutions of the Dyson-Schwinger equations (\ref{aeq}) and (\ref{beq}) 
have their asymptotic forms $A(s) \propto 1$ and $B(s) \propto 1/s$.
The low momentum strength of $g^2 D(p^2)$ is implied by 
the scale where the infrared
form is matched to the known asymptotic form.
From (\ref{eq49}) one can see that this matching scale,
which provides a natural cutoff scale for all the quark loop integrations, 
is of the order ${\cal{O}}\left ( \frac{\mu^2}{\epsilon} \right )$.

It should be mentioned in this context that the quarks obtained in that way are {\it confined} 
in the sense, that 
the quark propagator $G_0$ does not have a pole at timelike momenta.
This is a direct consequence of the $\frac{1}{q^4}$ singularity of the quark-quark interaction.

Finally we will have to perform integrations over the internal gluon momenta.
Using 
\begin{equation}
\int \frac{d^4 l}{(2\pi)^4} \frac{1}{l^{2n}} \frac{1}{(l+q)^{2m}} = 
\frac{1}{16\pi^2 } \frac{\Gamma(n+m-2)}{\Gamma(n)\Gamma(m)} B(2-m;2-n) (q^2)^{2-n-m}
\label{momint}
\end{equation}
we find that for $\epsilon\to 0$
\begin{eqnarray}
&& \int \frac{d^4 l}{(2\pi)^4}  \, g^2 D_s (l) 
 \, g^2 D_s (q+l) \, {\quad \propto \quad} \, \mu^2 \frac{\mu^2}{\epsilon} \frac{1}{q^4}
\nonumber \\
&& \int \frac{d^4 l}{(2\pi)^4}  l^2 \, g^2 D_s (l) 
 \, g^2 D_s (q+l) \, {\quad \propto \quad} \, \mu^2 \frac{\mu^2}{\epsilon} \frac{1}{q^2} 
\nonumber \\
&&\int \frac{d^4 l_1}{(2\pi)^4}   \frac{d^4 l_2}{(2\pi)^4}
\, g^2 D_s (l_1) 
\, g^2 D_s (l_1 + l_2) 
\, g^2 D_s (l_1 + l_2 + q ) 
\, {\quad \propto \quad} \, 
\mu^2 \left( \frac{\mu^2}{\epsilon} \right)^2  \frac{1}{q^4}
\nonumber \\
&&
\int \frac{d^4 l_1}{(2\pi)^4}   \frac{d^4 l_2}{(2\pi)^4} {l_1}^2 
\, g^2 D_s (l_1) 
\, g^2 D_s (l_1 + l_2) 
\, g^2 D_s (l_1 + l_2 + q ) 
\,
{\quad \propto \quad} \, 
\mu^2 \left( \frac{\mu^2}{\epsilon} \right)^2  \frac{1}{q^2}
\label{gluonint}
\end{eqnarray}
and so forth.

These observations provide us with the following {\it scaling rules} for each of the $\bar{q}q$
annihilation terms (\ref{eq312a},\ref{eq313},\ref{eq313a},...) in the limit $\epsilon\to 0$:
\begin{enumerate}
\item The scalar part of the quark self energy $B(s)$ scales as 
$\sqrt{\frac{\mu^2}{\epsilon}}$.
\item The vector part of the quark self energy $A(s)$ as well as all derivatives 
$A^\prime (s), B^\prime (s), \dots$ remain finite.
\item $X(s) = s A^2 (s) + B^2 (s) \propto \frac{\mu^2}{\epsilon}$.
\item Each quark loop integration $\int d^4 p$ gives a factor $\left ( \frac{\mu^2}{\epsilon} \right )^2$ 
and each additional quark momentum $p_\mu$ a factor $ \frac{\mu^2}{\epsilon}$.
\item The integration over $k$ internal gluon exchange lines gives a factor $\mu^2 
\left ( \frac{\mu^2}{\epsilon} \right )^{k-1}$.
\end{enumerate}

\subsection{Scaling Behavior of the $\bar{q}{q}$ Annihilation Diagrams; Generation of a 
Mass Term for the $\pi^0$ }
Let us now consider the limit $\epsilon \to0$ for each term in the series (\ref{eq312}) 
using the low energy expansion (\ref{gexp}) for $G_U$.
Hereby we will only consider terms in which each of the quark loop has at least one background 
field insertion.  
Diagrams with a quark loop which contains no background field insertion could be absorbed in a redefinition
of the gluon $n$ point function at NLO in $\frac{1}{N_c}$.
Using the explicit form (\ref{fourier}) for $G_1$ 
and making use of the scaling rules which were established in the last subsection leaves us with
the following leading order contributions of  
${\cal S}_{ann}$ quadratic in $\pi^0$:
\begin{enumerate}
\item The LO contribution from the 2 gluon diagram (Fig.\ref{fig3}(b)) is
\begin{eqnarray}
{\cal S}^{2g} \quad \propto \quad
&& 
\left [ \int ds s A(s) \left ( \frac{A(s)}{X(s)} \right ) 
\left ( \frac{B(s)}{X(s)} \right )^2 \right ]^2
\, \mu^2 \left ( \frac{\mu^2}{\epsilon} \right ) \, \cdot \,
\nonumber \\
&&
\int \frac{d^4 q}{(2\pi)^4}  
\int d^4 R_1 \pi^0 (R_1) 
\int d^4 R_2 \pi^0 (R_2) 
e^{- i q (R_1 - R_2)}
\label{eqs2gsc1}
\end{eqnarray}
which gives the total scaling behavior
\begin{equation}
{\cal S}^{2g} \quad \propto \quad \int 
\frac{d^4 q}{(2\pi)^4}  
\pi^0 (q) \pi^0 (-q) \mu^2 
 \left ( \frac{\mu^2}{\epsilon} \right )
\label{eqs2gsc2} \,
\end{equation}
where $\pi^0 (q)$ denotes the Fourier transform of $\pi^0 (x)$.
\item The LO contribution from a diagram with 
an odd number $2k+1 \ge 3$ of gluon exchanges 
(c.f. e.g.Fig.\ref{fig3}(c) for $2k+1 =3$) is
\begin{eqnarray}
{\cal S}^{(2k+1)g} \quad \propto \quad 
&&
\left [ \int ds s A(s) 
\left ( \frac{B(s)}{X(s)} \right )^{2k+2} \right ]^2
\, \mu^2 \left ( \frac{\mu^2}{\epsilon} \right )^{2k} \,  \cdot \,
\nonumber \\
&&
\int \frac{d^4 q}{(2\pi)^4}  
\int d^4 R_1 \pi^0 (R_1) 
\int d^4 R_2 \pi^0 (R_2) 
e^{- i q (R_1 - R_2)}
\frac{1}{q^2}
\, .
\label{eqs3gsc1}
\end{eqnarray}
Applying the scaling rules mentioned above 
leads a total LO scaling behavior of
\begin{equation}
{\cal S}^{(2k+1)g} \quad \propto \quad \int 
\frac{d^4 q}{(2\pi)^4}  
\pi^0 (q) \pi^0 (-q) \mu^2 
 \left ( \frac{\mu^2}{q^2} \right ) \,
\left ( \frac{\mu^2}{\epsilon} \right )^2 \, .
\label{eqs3gsc2}
\end{equation}
We see that the degree of divergence is $\frac{1}{\epsilon^2}$
independent of the number $2k+1$ of exchanged gluons. 
\item The LO contribution from a diagram with 
an even number $2k \ge 4$ of gluon exchanges 
(c.f. e.g.Fig.\ref{fig3}(d) for $2k =4$) is
\begin{eqnarray}
&&
{\cal S}^{(2k)g} \quad \propto \quad 
\left [ \int ds s B^{(j)}(s)
\left ( \frac{B(s)}{X(s)} \right )^{2k+1} \right ]^2
\, \mu^2 \left ( \frac{\mu^2}{\epsilon} \right )^{2k-1} \,  \cdot \,
\nonumber \\
&&
\int \frac{d^4 q}{(2\pi)^4}  
\int d^4 R_1 \pi^0 (R_1) 
\int d^4 R_2 \pi^0 (R_2) 
e^{- i q (R_1 - R_2)}
(q^2)^{(j-1)}  \, .
\label{eqs4gsc1}
\end{eqnarray}
Here $B^{(j)} (s)$ denotes the j th derivative of $B$.
This renders the total scaling behavior
\begin{equation}
{\cal S}^{2k} \quad \propto \quad \int 
\frac{d^4 q}{(2\pi)^4}  
\pi^0 (q) \pi^0 (-q) \mu^2 
F (q^2)
\left ( \frac{\mu^2}{\epsilon} \right )^2 \,  ,
\label{eqs4gsc2}
\end{equation}
whose degree of divergence is again $\frac{1}{\epsilon^2}$ and again 
independent on the number $2k$ of exchanged gluons.
The $F(q^2)$ stands for
a polynomial in $q^2$ with $F(0) \ne 0 $. 
\end{enumerate}

Therefore it follows that if $\epsilon \to 0$ we are left with a term
quadratic in $\pi^0$ of the form
\begin{equation}
{\cal S}_{ann} \quad \propto \quad \int \frac{d^4 q}{(2\pi)^4}
\pi^0 (q) \left [ F(q^2) q^2 + M^2 \right ] 
\pi^0 (-q) 
 \, \left ( \frac{\mu^2}{q^2} \right ) \,
\left ( \frac{\mu^2}{\epsilon} \right )^2
\label{Sann1}
\end{equation}
where $M^2 \neq 0$.
By rescaling the flavorsinglett field $\pi^0$ as
\begin{equation}
\pi^0 \to \pi^0 \sqrt{\frac{ \mu^2 }{q^2}} \, \frac{1}{\epsilon}
\label{rescale}
\end{equation}
one obtains finally
\begin{equation}
{\cal S}_{ann} \quad \propto \quad \int \frac{d^4 q}{(2\pi)^4}
\pi^0 (q) \left [ F(q^2) q^2 + M^2 \right ] 
\pi^0 (-q)      \, .
\label{Sann2}
\end{equation}
This means however that the $\pi^0$ has acquired a finite mass, which, in turn,
shows that the effective action ${\cal S}_{ann}$ is not $U_A (1)$ invariant. 
Important hereby is the fact that both terms in (\ref{Sann1}), namely the 
mass term $M^2$ as well as the wave function $F(q^2) q^2$ have the same 
degree of divergence in the limit $\epsilon \to 0$. 
They are both proportional to $\frac{1}{\epsilon^2}$. 
We can therefore absorb this divergence in a redefinition of the field $\pi^0$ by means of 
(\ref{rescale}).

It is important to notice that the mechanism described above
works only for the flavorsinglett particle $\pi^0$,
whereas the flavoroctet particles remain massless
Goldstone bosons due to the vanishing flavor trace over the quark loop.
Therefore the chiral $SU(3) \otimes SU(3) $ symmetry is still maintained.

\section{The Witten--di Vecchia--Veneziano Action}
Finally let us look at the general form of the action ${\cal S}_{ann}$ in the low energy limit,
if we use a gluon propagator with the Kogut-Susskind
$\frac{1}{q^4}$ singularity and take the IR limes as described in the last section.
So far we have shown that there appear terms quadratic in the flavorsinglet but not in the 
flavoroctet pseudoscalar fields.
The question is now, what happens with the non quadratic terms.
Because of parity the number of fields must be even, therefore the next
possible term must be of fourth order.
Typical diagrams are displayed in Fig.\ref{fig6}.
Applying the method described in the last section one finds easily that
those diagrams scale proportional to $\frac{1}{\epsilon}$
independent of the number of exchanged gluons. 
After rescaling the fields due to (\ref{rescale}) one finds that the contribution to
 ${\cal S}_{ann}$ with four $\pi^0$ fields is proportional to $\epsilon^3$.  
More general the contribution to
${\cal S}_{ann}$ with $(2j)$ $\pi^0$ fields 
is proportional to $\epsilon^{3(j-1)}$.
This means that all diagrams with 4 or more pseudoscalar field insertions
are suppressed by higher powers of  
$\epsilon $ compared to those with 2 pseudoscalar fields
and therefore
vanish in the limes $\epsilon \to 0$.
In other words the low energy limit of the
effective action coming from the ${\bar q}q$ annihilation diagrams
is of the form
\begin{equation}
{\cal S}_{ann} 
=
\lambda \, \int d^4 x \pi^0 (x) \pi^0 (x) \,  = \lambda \, 
\left [ \mbox{LnDet} \, U \right ]^2  
\label{eq52}
\end{equation}
where $\lambda$ is a finite numerical constant.

This, however, is exactly the form of an effective action 
containing a flavoroctet pseudoscalar which has been obtained Witten, 
di Vecchia and Veneziano \cite{witten,vecchia1,veneziano,vv}
from general $\frac{1}{N_c}$ counting arguments and 
its effect in effective chiral Lagrangians was studied
by various authors \cite{rosenzweig,nath}.
This means that the action we have obtained by integrating  
over the higher mass excitations with an IR singular 
gluon propagator from the Kogut-Susskind type is fully
consistent with the analysis of refs. \cite{witten,vecchia1,veneziano,vv}.

\section{Summary and Discussion}
Let us briefly summarize the most important issues of our analysis:
We are considering a model truncation of QCD (\ref{gcm}), which is based
on a very general form of a quark-quark interaction defined by a model
gluon propagator $D(q^2)$.
We have shown that in this approach
we are able to generate $\bar{q}q$ annihilation diagrams as a series
of multi gluon exchanges by integrating out the non would be Goldstone degrees
of freedom while treating the would be Goldstone degrees of freedom at the mean field level.  
Those diagrams are of one order $\frac{1}{N_c}$ higher than the leading term.
We then have examined the conditions for the creation of a finite mass term for the flavorsinglet
pseudoscalar ground state.
It turned out that it is sufficient that the gluon propagator $D(q^2)$
diverges as $\frac{1}{q^4}$ in the infrared ($q^2 \to 0$).
As a special case we have looked at a Kogut-Susskind type IR divergence
$D(q^2) \propto \frac{1}{q^{4-\epsilon}}$ where $\epsilon$ is an IR regulator.
This type of quark-quark interaction gives rise to confined constituent quarks.
The propagator for the flavorsinglet pseudoscalar develops a 
finite mass
${m_{\pi^0}}$ in the IR limes $\epsilon \to 0$.
The effective action generated by these $\bar{q}q$ annihilation diagrams 
is therefore not invariant under $U_A (1)$ transformations.
The $U_A (1)$ breaking is a consequence of the {\it highly nonlocal quark-quark
interaction} which is generated by $D(x-y)$ and the infrared singularities which are induced
by this nonlocality.
Taking the formally infinite functional traces while applying an appropriate IR regularization
prescription leaves one with a finite mass term for the flavorsinglet pseudoscalar channel.
The flavoroctet pseudoscalars remain massless Goldstone bosons and the chiral $SU(3) \otimes SU(3)$ symmetry remains 
untouched.
Moreover we have shown that a low energy expansion of the action action coming from these $\bar{q} q$ annihilation diagrams 
gives exactly the well known form $[\mbox{LnDet} U]^2$ proposed by Witten, di Vecchia and Veneziano
if the IR limes is taken as described above.
Therefore our approach is consistent with general $\frac{1}{N_c}$ counting.  

It is interesting to compare our case with the Nambu-Jona-Lasinio (NJL) type models \cite{njlreviews},
which, as we have mentioned earlier, can be regarded as a special case of the model truncation we consider,
namely where the quark-quark interaction reduces to a 
{\it local} four point interaction in coordinate space
$ D(x-y) \propto \delta^4 (x-y)$ and an appropriate UV cutoff is applied.
In momentum space $D(q^2)$ is then a finite constant and according to our analysis it is therefore
not possible to generate a finite $\pi^0$ mass in the NJL by this way.
On the other hand instanton based approaches \cite{dyakonov,zahed,shuryak},
which consider topological nontrivial vacuum gauge field configurations, are able to 
generate automatically an $U_A (1)$ breaking  
local ('t Hooft) interaction \cite{thooft},
which is a four quark interaction for $SU(2)$ flavor and a six quark interaction for
$SU(3)$ flavor.  
However, if one starts from a fully $U(3) \otimes U(3)$ invariant four quark interaction from the NJL type,
one 
is forced to add an $U_A (1)$ breaking but $SU(3)\otimes SU(3)$ invariant 
interaction 
by hand in order to obtain 
a finite mass for the flavorsinglet (see e.g. ref.\cite{dm}).
This can be either the 't Hooft term or the Witten - di Vecchia - Veneziano term, which we have discussed in the last section.
We want to point out once more that instantons are not present in the approach we have used, but all nonperturbative
effects are due to the IR behavior of the quark-quark interaction.
It seems, however, that instantons and our nonperturbative form for the quark-quark interaction
describe in different ways nonperturbative QCD, as it is reflected 
e.g in dynamical chiral symmetry breaking, 
the existence of nonperturbative vacuum condensates, or, as in our case, 
the $U_A (1)$ anomaly and $\eta'$ mass. 

It is clear that our treatment is far from being able to make a quantitative statement 
about the flavorsinglet pseudoscalar mass at the present stage.
In order to do so we would have to sum up the whole series of  multi gluon exchange diagrams
(\ref{eq312}),
which is technically more involved and will therefore be postponed to 
a separate analysis.

\acknowledgements
This work has been supported by the DOE (grant \# DE-FG06-90ER40561)
and the NSF (grant \# PHYS-9310124 and \# PHYS-9319641).
We would like to thank 
R. Alkofer (Universit\"at T\"ubingen),
L. von Smekal (Argonne National Laboratory),
P. Tandy (Kent State University) and 
V. Dmitra\v{s}inovi\'{c} (University of South Carolina) 
for useful discussions and comments.

\begin{figure}
\caption{Expansion of ${\cal S}_{ann}$
into multigluon exchange diagrams.
Each quark line is dressed and has an arbitrary number of chiral background field
insertions.
These diagrams are of order ${\cal O}(1)$, i.e. suppressed by one order
$\frac{1}{N_c}$ compared with the ones in Fig.{\protect{\ref{fig1}}} 
and Fig.{\protect{\ref{fig2}}}.  }   
\label{fig3}
\end{figure}

%

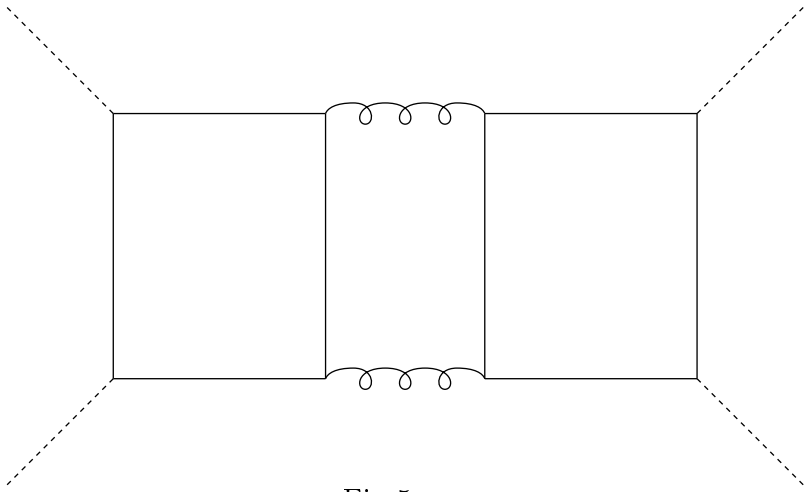
\begin{figure}
\caption{Typical diagram of fourth order in the pseudoscalar fields}
\label{fig6}
\end{figure}


\begin{references}

\bibitem{chipt}
J. Gasser and H. Leutwyler, Ann.Phys. (N.Y.) {\bf 158}, 142 
(1983); Nucl.Phys. B {\bf 250}, 465 (1985);
G. Ecker, Prog.Part.Nucl.Phys. {\bf 35}, 1 (1995), and references therein; 
V. Bernard, N. Kaiser and U. G. Meissner, Int.J.Mod.Phys. E {\bf 4}, 193 (1995),
and references therein.

\bibitem{witten}
E. Witten, Nucl.Phys. B {\bf 149}, 285 (1979); 
Nucl.Phys. B {\bf 156}, 269 (1979); 
Ann.Phys. (N.Y.){\bf 128}, 363 (1980). 

\bibitem{holstein}J. F. Donoghue, E. Golowich, and B. Holstein, 
{\it Dynamics of the Standard Model} (Cambridge University Press, 1992), and 
references therein.

\bibitem{ks}
J. Kogut and L. Susskind, Phys.Rev. D{\bf 10}, 3468 (1974).

\bibitem{marciano}
W. Marciano and H. Pagels, Phys. Rep.{\bf 36}, 137 (1978).


\bibitem{mandelstam}
S. Mandelstam, Phys.Rev. D{\bf 20}, 3223 (1979).

\bibitem{tuebingen}
A. Hauck, R. Alkofer and L. von Smekal, hep-ph/9604430. 

\bibitem{thooft}
G. 't Hooft, Phys.Rev. D{\bf 14}, 3432 (1976);
Phys.Rep.{\bf 142}, 357 (1986). 

\bibitem{shuryak}
T. Schaefer and E. Shuryak, hep-ph/9610451, and references therein. 

\bibitem{njl}
Y. Nambu and G. Jona-Lasinio, Phys.Rev.{\bf 122}, 345 (1961); 
{\bf 124}, 246 (1961).

\bibitem{njlreviews}
U. Vogl and W. Weise, Prog.Part.Nucl.Phys.{\bf 27}, 195 (1991); \\
S. Klevansky, Rev.Mod.Phys.{\bf 64}, 649 (1992); \\
T. Hatsuda and T. Kunihiro, Phys.Rep.{\bf 247}, 221 (1994); \\ 
R. Alkofer, H. Reinhardt and H. Weigel,  Phys.Rep.{\bf 265}, 139 (1996); \\
C. Christov et al., Prog.Part.Nucl.Phys.{\bf 37}, 1 (1996); \\
and references therein.


\bibitem{vecchia1}
P. di Vecchia, Phys.Lett. B{\bf 85}, 357 (1979).

\bibitem{veneziano}
G. Veneziano, Nucl.Phys. B{\bf 159}, 213 (1979).

\bibitem{vv}
P. di Vecchia and G. Veneziano, Nucl.Phys. B{\bf 171}, 273 (1980).  

\bibitem{rosenzweig}
C. Rosenzweig, J. Schechter and C. G. Trahern, 
Phys.Rev. D{\bf 21}, 3388 (1980).


\bibitem{nath}
R. Arnowitt and P. Nath, Nucl.Phys. B{\bf 209}, 234 (1980). 


\bibitem{gasser89}
G. Ecker, J Gasser, A. Pich, and E. de Rafael, Nucl. Phys. B{\bf 321}, 
311 (1989).


\bibitem{donoghue}
J. F. Donoghue, hep-ph/9506205, hep-ph/9607351.

\bibitem{frankmeissner}
M. R. Frank and T. Meissner,
Phys. Rev. C{\bf 53}, 2410 (1996). 

\bibitem{cahill1}
R. T. Cahill and C. D. Roberts, Phys.Rev. D{\bf 32}, 2419 (1985);
C. D. Roberts, R. T. Cahill  and J. Praschifka, Ann.Phys. (N.Y.) 
{\bf 188}, 20 (1988).

\bibitem{roberts}
C. D. Roberts, R. T. Cahill, M. E. Sevior and N. Iannella;
Phys.Rev. D{\bf 49}, 125 (1994).

\bibitem{gunner}
R. T. Cahill and S. Gunner, Phys.Lett. B{\bf 359}, 281 (1995).

\bibitem{mesons}
M. R. Frank and P. C. Tandy, Phys. Rev. C{\bf 49}, 478 (1994);
M. R. Frank, Phys. Rev. C{\bf 51}, 987 (1995);
M. R. Frank, K.L. Mitchell, C. D. Roberts and P.C. Tandy, Phys.Lett. B{\bf 359}, 17 (1995);
C. Roberts, Nucl. Phys. A{\bf 605},475 (1996);
D. Kekez and D. Klabucar, Phys.Lett. B{\bf 387}, 14 (1996);
C. Burden, L. Quian, C. Roberts, P. Tandy and M. Thomson,
Phys.Rev. C {\bf 55}, 2649 (1997).

\bibitem{soliton}
M. R. Frank, P. C. Tandy and G. Fai, Phys. Rev. C{\bf 43}, 2808 (1991); 
P. C. Tandy and M. R. Frank, Aust.J.Phys.{\bf 44}, 181 (1991); 
M. R. Frank and P. C. Tandy, Phys.Rev. C{\bf 46}, 338 (1992);
C. W. Johnson, G. Fai and M. R. Frank, Phys.Lett. B{\bf 386}, 75 (1996).

\bibitem{cahill2}
R. T. Cahill, Nucl.Phys. A{\bf 543}, 63 (1992).

\bibitem{meissner}
T. Meissner, Phys.Lett B{\bf 405}, 8 (1997).

\bibitem{km}
L.S. Kisslinger and T. Meissner, hep-ph/9706423, subm. to Phys.Rev. C. 


\bibitem{robertswilliams}
C. D. Roberts and A. G. Williams, Prog.Part.Nucl.Phys.{\bf 33}, 477 (1994).

\bibitem{tandyreviews}
P. C. Tandy, 
nucl-th/9705018, Prog.Part.Nucl.Phys{\bf 39} 1997 to appear.


\bibitem{kleinert}
H. Kleinert, Phys. Lett. B{\bf 62}, 429 (1976); 
in {\it Proc. of the 1976 School of Subnuclear Physics}, Erice (Plenum, 1978); 
Fortschr.Phys.{\bf 30}, 351 (1982).


\bibitem{shrauner}
E. Shrauner, Phys.Rev. D{\bf 16}, 1887 (1977).

\bibitem{dyakonov}
D.I. Dyakonov and V.Yu. Petrov, Nucl.Phys. B{\bf 245}, 259 (1984);
Nucl.Phys. B{\bf 272}, 457 (1986).

\bibitem{zahed}
M.A. Nowak, J.J.M. Verbaarschot and I. Zahed, Phys.Lett. B{\bf 228}, 251 (1989).

\bibitem{dm}
V. Dmitra\v{s}inovi\'{c}, Phys.Rev. C{\bf53}, 1383 (1996);
Phys.Rev. D{\bf 56}, 247 (1997). 




\end{references}
\end{document}